\documentclass[sigconf]{acmart}
\AtBeginDocument{%
  \providecommand\BibTeX{{%
    \normalfont B\kern-0.5em{\scshape i\kern-0.25em b}\kern-0.8em\TeX}}}

\setcopyright{CC}
\copyrightyear{2024}
\acmYear{2024}
\acmDOI{}

\acmBooktitle{CHI '24 Workshop: LLMs as Research Tools: Applications and Evaluations in HCI Data Work, May 11--16, 2024, Honolulu, Hawai'i} 

\begin{document}

\title{A Framework For Discussing LLMs as Tools for Qualitative Analysis}
\author{James Eschrich}
\email{jamesae2@illinios.edu}
\orcid{0009-0009-7608-3862}
\affiliation{%
  \institution{University of Illinios, Urbana-Champaign}
  \streetaddress{201 North Goodwin Avenue}
  \city{Urbana}
  \state{Illinios}
  \country{USA}
  \postcode{61801-2302}
}
\author{Sarah Sterman}
\email{ssterman@illinois.edu}
\affiliation{%
  \institution{University of Illinios, Urbana-Champaign}
  \streetaddress{201 North Goodwin Avenue}
  \city{Urbana}
  \state{Illinios}
  \country{USA}
  \postcode{61801-2302}
}

\renewcommand{\shortauthors}{Eschrich and Sterman}

\begin{abstract}
We review discourses about the philosophy of science in qualitative research and evidence from cognitive linguistics in order to ground a framework for discussing the use of Large Language Models (LLMs) to support the qualitative analysis process. This framework involves asking two key questions: ``is the LLM \textbf{proposing} or \textbf{refuting} a qualitative model?'' and ``is the human researcher \textbf{checking} the LLM's decision-making directly?''. We then discuss an implication of this framework: that using LLMs to \textit{surface counter-examples for human review} represents a promising space for the adoption of LLMs into the qualitative research process. This space is promising because it is a site of overlap between researchers working from a variety of philosophical assumptions, enabling productive cross-paradigm collaboration on tools and practices. \end{abstract}




\maketitle

\section{Introduction}
LLMs can perform better than traditional NLP techniques at certain "natural language understanding" tasks like "miscellaneous text classification" \cite{yangHarnessingPowerLLMs2024}, which raises the possibility that they could potentially be used for qualitative data analysis. Emergent literature investigates some possibilities such as using LLMs for deductive coding \cite{xiaoSupportingQualitativeAnalysis2023, taiExaminationUseLarge2024} and inductive thematic analysis \cite{depaoliPerformingInductiveThematic2023}. However, there is disagreement about whether, where, and to what extent LLMs should or could be used to support the qualitative analysis process \cite{robertsArtificialIntelligenceQualitative2024, feustonPuttingToolsTheir2021}.

In order to structure discussion about whether and under what conditions LLMs could be used to support the qualitative analysis process, we review relevant background on the philosophy of science, discussing \textbf{positivism} and \textbf{constructivism} using \cite{haylesConstrainedConstructivismLocating1993}'s language of \textit{consistency} and \textit{congruence}. We also review evidence from cognitive linguistics about how humans understand language. Then, based on this background, we propose a framework for structuring discussions around the use of LLMs to support qualitative analysis. This framework involves asking ``is the LLM \textbf{proposing} or \textbf{refuting} a qualitative model?'' and ``is the human researcher \textbf{checking} the LLM's decision making directly?'' Finally, we will discuss an implication of this framework: that \textit{using LLMs to surface counter-examples for human review} from a qualitative dataset represents a promising space for the adoption of LLMs into the qualitative analysis process.
\section{Background}
We characterize two positions on the philosophy of science, \textbf{positivism} and \textbf{constructivism}, using \cite{haylesConstrainedConstructivismLocating1993}'s language of \textit{consistency} and \textit{congruence}. We then review evidence from cognitive linguistics about how humans process language, and discuss the implications for using LLMs in qualitative analysis from both \textbf{positivist} and \textbf{constructivist} perspectives.

\subsection{Positivism and Constructivism}
Discourse about the philosophy of science in qualitative research has a long and complex history \cite{brinkmannHistoricalOverviewQualitative2014, denzinIntroductionDisciplinePractice2005}. A variety of "ways of knowing" \cite{kelloggEpilogue2014} are taken up by researchers, who are often conceptualized as operating within "paradigms"—multiple perspectives on foundational philosophical issues packaged into coherent systems \cite{gubaAlternativeParadigmDialog1990}. This picture is over-simplified \cite{denzinIntroductionDisciplinePractice2017} and potentially limiting \cite{chafeRejectingChoicesProblematic2023}, but we will further limit our discussion to two paradigms—\textbf{positivism} and \textbf{constructivism}—in reference to which many other paradigms, like \textbf{post-positivism} or \cite{haylesConstrainedConstructivismLocating1993}'s ``constrained constructivism'' are defined. We characterize \textbf{positivism} and \textbf{constructivism} using \cite{haylesConstrainedConstructivismLocating1993}'s language of \textit{congruence} and \textit{consistency}, which we find more useful here than \cite{gubaAlternativeParadigmDialog1990, chamberlainMethodolatryQualitativeHealth2000}'s language of ontology, epistemology, and methodology \footnote{This is in part because some \textbf{constructivist} perspectives reject "the ontology/epistemology distinction" \cite{gubaAlternativeParadigmDialog1990}.}.

\textbf{Positivism} is oriented towards identifying and validating \textit{congruent} models—models for which a ``one-to-one correspondence'' \cite{haylesConstrainedConstructivismLocating1993} can be established between the structure of the model and the structure of external reality. Since models which are \textit{inconsistent} with observation cannot be \textit{congruent}, discriminating between \textit{consistent} and \textit{inconsistent} models is relevant. However, neither the question of how to discriminate between different \textit{consistent} models or the possible existence of \textit{unknown} or unknowable models are of particular concern. This is because if \textit{congruence} with reality is established, the possibility spaces of the \textit{consistent} and the \textit{unknown} become irrelevant.

\textbf{Constructivism} denies the possibility of establishing \textit{congruent} models, instead radically acknowledging induction's inherent limits \cite{gubaAlternativeParadigmDialog1990} and the \textit{unknown}—models not yet tested, not yet conceived, or fundementally unknowable by humans \cite{haylesConstrainedConstructivismLocating1993}. \textbf{Constructivism} thus reverses the emphases of \textbf{positivism}, focusing on exploring the possibility space of \textit{consistent} models while engaging reflexively with the limits of our perspectives and, thus, the omnipresent \textit{unknown}.

The language of \textit{consistency} and \textit{congruence} helps reveal these paradigms as two points on a continuum; different positions can be located by examining how much attention is paid to \textit{congruent} models over \textit{unknown} models or \textit{consistent} models over \textit{inconsistent} models. This language also highlights an important area of overlap between \textbf{constructivism} and \textbf{positivism}. Both paradigms are interested in rejecting models that are \textit{inconsistent} with observation: the \textbf{positivist}\footnote{In this paper, `the positivist' and `the constructivist' represent "ideal ... and extreme" \cite[p.~5]{chafeRejectingChoicesProblematic2023} characterizations of the research paradigms deployed for explanatory purposes, rather than individuals who may use or identify with a paradigm for specific reasons within specific contexts.} while trying to establishing \textit{congruence} with reality and the \textbf{constructivist} while balancing an exploration of different \textit{consistent} models with a continual deference to the \textit{unknown}.  

\subsection{LLMs and The Implications of Cognitive Linguistics}
Evidence increasingly suggests that the human ability to understand natural language is based on both cultural and embodied experience \cite{hampeEmbodimentDiscourseDimensions2017, gibbsMetaphorGroundedEmbodied2004}. This poses a strong challenge to any attempts to use LLMs in the qualitative analysis process. However, the nature of this challenge varies across philosophical frameworks. For \textbf{positivists}, concerned with discovering and validating models congruent with reality, the process by which LLMs produce claims is not intrinsically important; the structure of the reality we are trying to model exists independently and can theoretically be arrived at through multiple means. However, attempting to measure how consistently LLMs arrive at congruence with reality remains important. 

For \textbf{constructivists}, concerned with discriminating between consistent models and keeping the unknown in view, the process by which LLMs make claims about language data is our primary concern. Constructivists understand knowledge as enabled by "particular sets of sensory apparatus located within specific cultures and times" \cite{haylesConstrainedConstructivismLocating1993}; thus, the constructivist's ability to evaluate claims produced by an LLM requires understanding how the LLM came to produce those claims. What value, if any, claims originating from LLMs have from this perspective is uncertain and unclear.

\section{A Framework for Discussing LLMs for Qualitative Analysis}
With the philosophical and linguistic background reviewed, we propose a framework (Table~\ref{tab:framework}) for discussing the use of LLMs for qualitative analysis. This framework is structured around two questions. The first question, ``is the LLM \textbf{proposing} or \textbf{refuting} a qualitative model?'' highlights the important "asymmetry" \cite{haylesConstrainedConstructivismLocating1993} between affirming consistency and refuting consistency. The second question, ``is the human researcher \textbf{checking} the LLM's decision-making directly?'' draws attention to the complex processes by which humans come to understand language and to what extent LLMs are capable of either supporting or replacing those processes.
\begin{table*}[ht]
\centering
  \caption{A Framework for Discussing LLMs for Qualitaive Analysis}
  \label{tab:framework}
\begin{tabular}{p{0.2\linewidth} | p{0.4\linewidth} | p{0.4\linewidth}}
\hline
  & LLM is \textbf{proposing} a qualitative model (\textit{e.g.} codes, themes, categories, clusters) & LLM is \textbf{refuting} a qualitative model (\textit{e.g.} counter-examples) \\
\hline

Human researcher is \textbf{checking} the LLM's decisions directly & \textit{Example}: clustering related codes for human review, automated/suggested open coding with human review\newline\newline\textit{Positivist}: How is this supporting or hindering my ability to validate congruence with reality?\newline\newline\textit{Constructivist}: Do I agree the LLM's proposed model is consistent with these observations? How is this supporting or hindering my ability to explore other consistent models? How is this affecting my ability to reflect on my experience?& \textit{Example}: suggesting counter-examples to a developing theory/category/theme\newline\newline\textit{Positivist}: Should this model be rejected as inconsistent? Can I account for this counter-example?\newline\newline\textit{Constructivist}: Should this model be rejected as inconsistent? Can I account for this counter-example? \\
\hline
Human researcher is \textbf{NOT checking} the LLM's decisions directly & \textit{Example}: automated open coding without human review\newline\newline\textit{Positivist}: How can I be confident that the LLM’s claims are congruent with reality? If they aren’t, are they similar enough to an “average human’s” claims to be useful?\newline\newline \textit{Constructivist}: How can we evaluate or use a qualitative model that does not arise from an inter-subjective human experience? Should we be doing this in the first place? & \textit{Example}: deductive coding without human review (i.e. for each observation, eliminating codes which are inconsistent)\newline\newline\textit{Positivist}: How well does an LLM perform compared to a human in correctly identifying data inconsistent with a given model?\newline\newline\textit{Constructivist}: How do the inconsistencies an LLM identifies compare with mine? Other people's? Is the LLM biased in a certain way? How is this shaping my analysis? Should we be doing this in the first place? \\
\hline
\end{tabular}
\end{table*}

\section{Discussion}
The upper-right quadrant of Table~\ref{tab:framework}, which we will call `Using LLMs to \textit{surface counter-examples for human review}' is unique in that, in this case, the positivists and constructivists have the same questions. This is because both positivists and constructivists are interested in discarding models which are inconsistent with observation, although they may differ on what sort of evidence implies inconsistency and in what contexts that inconsistency holds. 

We can also articulate this point using the language of ontology, epistemology, and methodology. Negating a model is ontologically neutral; `how can we judge whether this model is inconsistent with observation?' is a purely epistemological question. Moreover, when a human researcher is the one answering that question, using an LLM to support the process is not \textit{inherently} problematic epistemologically; no tool for performing qualitative analysis is neutral, and all tools influence the process in some way \cite{gilbertToolsAnalyzingQualitative2014}. Thus, only the methodological question remains: is using LLMs to \textit{surface counter-examples for human review} useful/effective for doing qualitative analysis?

One way using LLMs to \textit{surface counter-examples for human review} could be useful is by enabling researchers to work with very large datasets. Existing mixed-methods work has used topic modeling to ``purposively sample'' from very large datasets like posts on a subreddit \cite{gauthierWillNotDrink2022}. This is an exciting approach; however, topic modeling represents a situation in which a language model is `\textbf{proposing} a qualitative model'. As an alternate approach, an initial sample could be selected for a preliminary qualitative analysis using traditional methods like random or stratified sampling. This preliminary analysis could then be refined by using an LLM to \textit{surface counter-examples for human review} from the much larger dataset. This somewhat resembles the use of theoretical sampling in grounded theory to ``develop and refine'' emerging categories \cite[p.~205]{charmazConstructingGroundedTheory2014}. 

How using LLMs to \textit{surface counter-examples for human review} might positively or negatively affect the results of a qualitative analysis is an open question. However, restricting the LLM to \textit{surfacing counter-examples for human review} ensures that (a) all claims about which qualitative models are consistent with observation originate from human researchers and (b) no qualitative model is rejected by an LLM-identified observation without giving human researchers a chance to revise the model to account for it. This creates a point of compatibility which could enable tools and practices to be collaboratively developed and productively employed across philosophical boundaries.
\section{Conclusion}
Based on a review of the philosophy of science in qualitative research and evidence from cognitive linguistics, we propose a framework for discussing the use of LLMs for qualitative analysis. This framework is based on two key questions: ``is the LLM \textbf{proposing} or \textbf{refuting} a qualitative model?'' and ``is the human researcher \textbf{checking} the LLM's decision-making directly?''. We discuss the potential of using LLMs to \textit{surface counter-examples for human review}, which this framework identifies as promising because it is a site of overlap between different philosophical perspectives. We hope our framework and discussion will help researchers working from different research paradigms discuss the potential of LLMs as tools for supporting qualitative analysis.

\begin{acks}
We would like to thank: Camille Cobb for her advice and perspective; Karen Wickett and Melissa Ocepek as well as Bhavana Bheem and the rest of IS 590 for their support and feedback; and Steven Eschrich, Ethan Eschrich, Sarah Eschrich, and Suzanne Eschrich for their feedback, advice, and support.
\end{acks}

\bibliographystyle{ACM-Reference-Format}
\bibliography{sample-base}

\end{document}